\documentstyle[manuscript,aps]{revtex}
\begin{document}
\title{Modeling and manufacturability assessment of
bistable quantum-dot cells}
\author {M. Governale, M. Macucci, G. Iannaccone, C. Ungarelli}
\address{Dipartimento di Ingegneria dell'Informazione,
Universit\`a di Pisa\\
Via Diotisalvi, 2, I-56126 Pisa, Italy}
\author {J. Martorell}
\address{Dept. d'Estructura i Constituents de la Materia, Facultat 
de F\'{\i}sica,\\
Universitat de Barcelona, Barcelona 08028, Spain}
\date{\today}
\maketitle
\begin{abstract}
We have investigated the behavior of bistable cells made up of four
quantum dots and occupied by two electrons, in the presence of realistic
confinement potentials produced by depletion gates on top of a GaAs/AlGaAs
heterostructure. Such a cell represents the basic building block for
logic architectures based on the concept of Quantum Cellular Automata (QCA) and
of ground state computation, which have been proposed as an alternative to 
traditional transistor-based logic circuits. We have focused 
on the robustness of the operation of such cells with respect to asymmetries 
deriving from fabrication tolerances. We have developed
a 2-D model for the calculation of the electron density in a driven cell 
in response to the polarization state of a driver cell. Our method is
based on the one-shot Configuration-Interaction technique, adapted from
molecular chemistry. From the results of our simulations, we conclude 
that an implementation of QCA logic based on simple ``hole-arrays'' is
not feasible, because of the extreme sensitivity to fabrication tolerances.
As an alternative, we propose cells defined by multiple gates, where geometrical
asymmetries can be compensated for by adjusting the bias voltages. Even though
not immediately applicable to the implementation of logic gates and not
suitable for large scale integration, the proposed cell layout should allow 
an experimental demonstration of a chain of QCA cells.
\end{abstract}

\section{Introduction}
Several proposals for the implementation of logic functions and data
processing based on the concept of Quantum Cellular
Automata (QCA) and ground state computation have 
appeared\cite{Toglog,Lentapp,Dash,Bandy} in
the literature recently. Tougaw {\sl et al.}\cite{Toglog} 
devised
an architecture (commonly known as ``Notre Dame architecture'') based on
bistable cells which couple electrostatically to their nearest neighbors.
Each cell consists of four (or five) quantum dots and contains a total of
two electrons.
In the absence of external electric fields, the electrons occupy each dot
with equal probability. In the presence of a nearby (driver) cell with 
the two electrons forced in the dots along one of the two diagonals, alignment 
along the same diagonal will occur in the driver cell, in the hypothesis of  
potential barriers large enough as to localize the electrons.
Based on this principle, it is
possible to conceive of bistable cell arrays, in which the polarization
state enforced at the inputs, at the edges of the arrays, propagates in a
``domino'' fashion\cite{Lentwire}, until the ground state is reached
throughout the system, and the results of the computation are available
in the form of the polarization state of the output cells, also
located at the edges of the arrays.

Many issues must be resolved before this computational paradigm can be
implemented in practice: non-invasive detectors are needed to probe the
polarization state of the outputs without perturbing the ground state of the
system; provision must be made for a time evolution of the system
that is both fast and reliable; design solutions for the basic
cell must be developed, yielding a reasonable robustness to fabrication
tolerances and compatible with large-scale integration on a single chip.

The focus of the work we are presenting is specifically on the
robustness of a single cell, coupled to a driver cell, to fabrication
tolerances and to asymmetries caused by fluctuations in the bias voltages 
applied
to the electrodes defining the cell. In particular, we have studied the effect
of geometrical and electrical asymmetries on the behavior of QCA cells defined
by means of lateral electrostatic confinement in a GaAs/AlGaAs heterostructure.

Although often neglected, robustness to fabrication tolerances and 
manufacturability are central problems affecting all proposed nanoelectronic 
devices\cite{Landauer}, and their solution is a prerequisite for any 
successful new technology.

We have considered a basic QCA cell with four quantum dots defined by realistic
two-dimensional
confinement potentials, which are computed from the shape of the metal gates
at the surface of the heterostructure and the voltages applied to them.
Calculation of the electron density in such a 2-D artificial molecule is a
challenging task. Iterative self-consistent methods fail to converge, due to
the relatively large electrostatic interaction and to the particular symmetries
associated with the problem.  For this reason, we have developed a
non-iterative technique based on the Configuration-Interaction method used in
molecular chemistry\cite{Lverde}. The main drawback of this technique is
that it requires rather large computational resources, and, thus,
we are presenting numerical results only for the case of occupancy of two
electrons per cell. (Work is currently in progress for the inclusion of up to
six electrons per cell.)

With respect to the approaches existing in the literature\cite{Togsat},
our method allows a direct quantitative estimate of the effects of
fabrication and bias tolerances on cell operation and does not require the
introduction of phenomenological parameters such as the tunneling energy,
which may be hard to evaluate with a realistic potential.

In the next section, we provide a detailed statement of the problem we intend
to solve and describe the cell model together with the technique we have
used for the computation of the 2D confinement potential. In the third
section, the solution of the many-body problem is discussed and the one-shot
Configuration-Interaction method is introduced. Numerical results for a cell
occupancy of two electrons and various types of asymmetries are presented in
Sec.~IV, where cell design criteria are also established.

\section{Statement of the problem and cell model}

Our aim is to investigate the behavior of two coupled QCA cells, each
of which is formed by four quantum dots and contains two electrons, as 
illustrated in Fig.~\ref{fig1}. 
Tunneling between the dots of the same cell is allowed, but not
between dots belonging to different cells. If the barriers separating
the dots within a cell are opaque enough, the electron wave functions will be
localized, and we shall observe a quasi-classical behavior: the two electrons
will repel each other and localize in two dots along a diagonal, so 
as to minimize the electrostatic energy. In the case of an isolated, symmetric
cell, alignment along either diagonal will occur with equal probability.
If another 
(driver) cell is placed in proximity to the cell we are investigating
(driven cell), as in the case represented in Fig.~\ref{fig1}, and the 
electrons in the
driver cell are taken to be aligned along a given diagonal, their electric
field will destroy the symmetry of the driven cell, lifting
the degeneracy between the two configurations. This will result in the
electrons in the driven cell lining up parallel to the
electrons in the driver cell.

Following Lent {\sl et al.}\cite{Lentapp} we define a cell polarization $P$ as

\begin{equation}
P \equiv {{(Q_1+Q_3) - (Q_2 +
Q_4)}\over{Q_1+Q_2+Q_3+Q_4}}\,,\label{prima}
\end{equation}
where $Q_i$ is the integral of the electron density $\rho(\vec{r})$ over
the $i-$th quadrant of a cell.
We divide each cell into four quadrants (see Fig.~\ref{fig2}) and
number them counterclockwise. 
The denominator of Eq.~(\ref{prima}) is the total number 
of electrons in the cell and therefore a constant. The procedure 
for calculating the electron density from the many-electron 
wave functions will be discussed in the following section. 

If the two electrons are aligned along the diagonal corresponding to the first
and third quadrant, $P=1$, while if they are aligned along the other diagonal,
$P=-1$. For finite height barriers, values of $P$ intermediate
between -1 and 1 are also possible. We define the cell-to-cell response
function as the function relating the polarization of the driven cell to
that of the driver cell. In order to compute the polarization of the driven
cell in response to each value of the polarization of the driver cell, we need
to solve for the electronic structure in the driven cell in the presence of
the electrostatic potential due to the driver cell.

For the driven cell we consider a two-dimensional model in the effective mass
approximation. Such a model is valid as long as the thickness of the 
dots in the
vertical direction, corresponding to the thickness of the 2-dimensional
electron gas (2DEG) from which they are obtained by lateral confinement, is
small compared to their other dimensions. The 2DEG is obtained by
modulation doping next to a GaAs/AlGaAs heterointerface.

The two-dimensional confinement potential in the plane of the 2DEG is obtained
as a result of the action of the metal gates, following the method proposed by
Davies {\sl et al.}\cite{D88,DLS95}, without including, to keep the problem
manageable from a computational point of view, the self-consistent
rearrangement of mobile charge within the heterostructure, except for that of
the two electrons confined in the cell. In other words, the potential due to
the gates is evaluated with the analytical expressions described below
and is used as the bare confinement potential for the
definition of a many-body Hamiltonian, whose ground state is then evaluated
with the Configuration-Interaction method. The occupancy of the cell is
fixed and corresponds to two electrons for all the numerical
results we shall be presenting. We assume Fermi level pinning at the surface
of the semiconductor, so that the electron-electron interaction can be treated
by the method of images\cite{MS94}, without requiring the solution
of the Poisson equation.

We have considered two basic gate configurations for the definition 
of the four quantum dots that make up a cell. The first 
configuration, represented in Fig.~\ref{fig3}, is rather simple and 
straightforward: the four dots are a consequence of 
four circular holes in a depleting gate covering the surface of 
the heterostructure\cite{Porochen}. The second configuration we have 
studied is described in Fig.~\ref{fig4} and is more complex: the four 
dots are defined by a set of seven metal gates which create four minima 
in the 2-D potential at the 2DEG level. 

The contribution to the bare confinement potential from each gate is computed
following the method developed in Refs. \cite{D88} and \cite{DLS95}.
The starting point is the
 well known result that gives the potential inside the semiconductor in terms
 of its boundary values at the plane of the  surface:
\begin{equation}
V_g(x,y,z) = {1\over {2\pi}}\int_S {{|z| V_g(x',y',0)}\over{(x- x')^2
+(y-y')^2+z^2}}  \ dx' \ dy' \ ,
\label{eq:pot1}
\end{equation}
where $z$ is the vertical coordinate, orthogonal to the heterostructure
layers, and the integration is performed over the gated surface $S$.
Given the applied voltages and the shapes of the gates, the
confining potential can be easily computed. For simple shapes one can derive
more compact expressions, by performing some of the integrals in
Eq.~(\ref{eq:pot1}) analytically. For the cases considered here, we have used
Eq.~(3.17) of \cite{D88} for gates with circular holes, and the equations in
Sect. III and Sect. IV of \cite{DLS95} for polygonal gates.

An example of the results obtained by this procedure is reported in 
Fig.~\ref{fig5}:
we show the confinement potential at a depth of 50~nm, produced by four holes
with a diameter of 90~nm, with a distance between the centers of 110~nm.
The gate voltage has been set at $-0.5$~V, in order to obtain interdot 
barriers of reasonable height.

Following Tougaw {\it et al.}\cite{Togsat}, a uniformly distributed positive 
background charge has been added to each cell. Such a charge does not alter 
the electrostatic energy splitting between the two cell configurations and 
plays a role, from the point of view of the cell-to-cell response function, 
only if the two cells are very close to each other. In this case, it helps
prevent alterations of the ground state of the driven cell due to 
the combined electrostatic repulsion of the two electrons in the driver 
cell, which would tend to push both electrons into the two dots
on the side further from the driver cell itself. 
In our model, this positive background would not actually 
be needed to achieve charge neutrality\cite{Togsat}, which is already ensured
by the presence of the gates and of positive charges in the donor layer.

\section{Configuration-Interaction Method}

As already mentioned, the solution of a many-electron problem in a 
potential such as that present in a QCA cell is rather challenging. 
Approaches that are typically used for the simulation of quantum 
dots, based on mean-field approximations of the potential seen by 
each electron and on iterative 
procedures\cite{Kumar,Stopa1,Stopa2,Mac1,Mac2}, fail to converge when 
applied to the four-dot cell. We can understand the reason for this 
failure considering that convergence of the self-consistent iterative 
procedures in this class of problems is more and more difficult to 
achieve as the electrostatic interaction increases\cite{Mac3}, and in 
the presence of quasi-degenerate states. As long as the electrostatic 
interaction is small compared to the confinement energy, it is just a 
perturbation of the latter, and iterative self-consistent procedures 
converge monotonically to the solution. Otherwise, underrelaxation 
techniques need to be used, but they also often fail when closely 
spaced states are present, due to symmetries or quasi-symmetries in 
the potential landscape: in this situation the charge will bounce 
back and forth between quasi-degenerate states in consecutive 
iterations, and convergence is never achieved. Since our calculations 
are currently performed at zero temperature, in order to compare with 
experimental results that are typically obtained in the 
tens of millikelvins range, techniques such as the Newton method, 
which are rather successful in the solution of coupled 
Schr\"odinger-Poisson problems at finite temperature, cannot be 
successfully used, due to the sharpness of the Fermi function at low 
temperatures. 

The technique we have implemented is based upon an approach often 
used in molecular chemistry\cite{Lverde}, the 
Configuration-Interaction (CI) method. It consists in approximating 
the $N$-electron wave function by a finite linear combination of Slater 
determinants. Most importantly, the CI method is a one-shot method, 
i.e. it does not involve an iterative calculation of the wave 
functions, and, hence, does not suffer from the previously described 
convergence/oscillation problems. 

In order to illustrate the CI method, let us consider an
N-electron non-relativistic Hamiltonian with a generic
two-body interaction $g(\vec{r}_i,\vec{r}_j)$:
\begin{eqnarray}
&&\hat{H}=\hat{H}_1+\hat{H}_2\,,
\label{Htot}\\
&&\hat{H}_1=\sum_{i=1}^{N}(-\frac{\hbar^2}{2\,m}\,\nabla^2_i + V(\vec{r}_i))
=\sum_{i=1}^{N} {h}(\vec{r}_i)\,,\\
&&\hat{H}_2=\sum_{i\,<\,j}g(\vec{r}_i,\vec{r}_j)\,,\quad{\rm with}\quad
g(\vec{r}_i,\vec{r}_j)=g(\vec{r}_j,\vec{r}_i)\,,
\end{eqnarray}
where $\hbar$ is the reduced Planck constant and $m$ is the effective 
mass of the electron (we consider the case of Gallium Arsenide, with
$m=0.067 m_0$, $m_0$ being the free electron mass).

As we are concerned with confined systems, we can consider a 
numerable complete basis $\{\varphi_i({\bf q})\}$, where ${\bf q} = 
(\vec{r},s)$ includes both spatial and spin coordinates, over 
which the single-particle wave function can be expanded . In the 
following, we shall refer to the $\varphi_i$'s as spin-orbitals. 
Using this basis, we build all the possible independent Slater 
determinants: 

\begin{equation}
\Phi_k=\frac{1}{\sqrt{N!}}\, \left | \begin{array}{cccc}
\varphi_{n_{1k}}({\bf q}_1) &\varphi_{n_{2k}}({\bf q}_1)
& \dots &\varphi_{n_{Nk}}({\bf q}_1) \\
\varphi_{n_{1k}}({\bf q}_2) &\varphi_{n_{2k}}({\bf q}_2)
& \dots &\varphi_{n_{Nk}}({\bf q}_2) \\
\dots & \dots & \dots & \dots \\
\varphi_{n_{1k}}({\bf q}_N) &\varphi_{n_{2k}}({\bf q}_N)
& \dots &\varphi_{n_{Nk}}({\bf q}_N)
\end{array}
\right | \,,
\end{equation}
where the index $k$ labels the Slater determinants and the integer
$n_{jk}$ specifies which spin-orbital appears in the $j-$th column of
the $k-$th Slater determinant.

The infinite set $\{\Phi_k\}$ is a complete orthonormal basis  for the
$N-$electron eigenfunctions of the Hamiltonian in Eq.~(\ref{Htot})
\cite{Slater}; the $i$-th eigenfunctions $\Psi_i$ can therefore be
written as 
\begin{equation}
\Psi_i=\sum_{k=1}^{\infty}c_{i\,k}\,\Phi_k\,\,.
\label{wf}
\end{equation}
To find the eigenfunctions of ${\hat H}$, we must solve the
secular equation

\begin{equation}
{\cal H}{\bf c}_i=E_i\,{\bf c}_i\,,
\label{secul}
\end{equation}
where  the infinite-dimensional  ``Hamiltonian matrix'' is
\begin{equation}
{\cal H}_{k\,k^{\prime}}=\langle\Phi_k|\hat{H}|\Phi_{k^{\prime}}\rangle=
\int\,\Phi_k^{*}(\hat{H}_1+\hat{H}_2)
\Phi_{k^{\prime}}\,\prod_{i=1}^{N}d{\bf q}_i \,
,\end{equation}
$E_i$ is the $i$-th eigenvalue of ${\cal H}$, the vector ${\bf
c}_i$ is made up of the coefficients $c_{i\,k}$
and $\int\,d{\bf q}_i\,$ stands for integration over the $i$-th spatial
coordinate {\it and} summation over spin orientations.

In practice, this approach cannot be implemented
``exactly'' (i.e. choosing a complete, infinite set of orthonormal
spin-orbitals). In our numerical work, we
consider a finite set of $M$ spin-orbitals $\{\varphi_j({\bf q})\}$, with
$j=1\dots\,M$.

Given $N$-electrons and $M$ spin-orbitals (with
$M\geq N$), it is possible to build ${\cal N}_{SD}$ different Slater
determinants, where
\begin{equation}
{\cal N}_{SD}=\left(\begin{array}{c}
M \\
N
\end{array}\right )\,.
\label{nsd}
\end{equation}
With this choice, the secular equation~(\ref{secul}) becomes
an ${\cal N}_{SD}\,\times\,{\cal N}_{SD}$ Hermitian eigenvalue
problem.

The number of nonzero matrix elements is less than 
$({\cal N}_{SD})^2$, because all the matrix elements between determinants
differing by more than two spin-orbitals do vanish, as a consequence of 
the selection rules (Slater's rules~\cite{Lverde}) presented in 
Appendix~A, where the general expressions for the elements of ${\cal H}$
are also provided.

The total number ${\cal M}_{NZ}$ of nonzero matrix elements is given by 
the following expression:
$$
{{\cal M}_{NZ}}=\left(\begin{array}{c}
M \\
N
\end{array}\right )\times
\left[1+
\left(\begin{array}{c}
M-N \\
1
\end{array}\right)\,
\left(\begin{array}{c}
N \\
1
\end{array}\right)
+
\left(\begin{array}{c}
M-N \\
2
\end{array}\right)\,
\left(\begin{array}{c}
N \\
2
\end{array}\right)
\right]
$$
\begin{equation}
=\left(\begin{array}{c}
M \\
N
\end{array}\right )\times
\left[1+(M-N)N +\frac{(M-N)(M-N-1)N(N-1)}{4}\right]
\label{geppo}
\end{equation}

Once the eigenvectors of Eq.~(\ref{secul}) have been obtained, the 
$N$-electron wave function $\Psi_i$ can be computed from Eq.~(\ref{wf}), 
and the corresponding electron density is simply given by:
\begin{equation}
\rho_i(\vec{r}_1)=N\,\sum_{s_1}\, \int
|\Psi_i(\vec{r}_1,s_1,{\bf q}_2,\dots,{\bf q}_N)|^2\,d{\bf q}_2
\dots d{\bf q}_N\,,
\label{eldens}
\end{equation}
where $s_1$ represents the spin orientation coordinate.

Since we are going to present numerical results for a cell occupancy of two 
electrons, 
we now focus our attention on the two-electron case. The 
Hamiltonian for the structure under study can be written:
\begin{equation}
\hat{H}=-\frac{\hbar^2}{2\,m}\,\nabla^2_1-\frac{\hbar^2}{2\,m}\,\nabla^2_2
+V_{\mbox{con}}(\vec{r}_1)+V_{\mbox{con}}(\vec{r}_2)+
V_{\mbox{driv}}(\vec{r}_1)+V_{\mbox{driv}}(\vec{r}_2)+
g(\vec{r}_1,\vec{r}_2)
\label{2elH}
\end{equation}
where $V_{\mbox{con}}$ is the confinement potential computed as
described in the previous section, $V_{\mbox{driv}}$ is the Coulomb
potential due to the charge distribution in the neighboring driver
cell, and $g(\vec{r}_1,\vec{r}_2)$ is the two-body interaction.
The two-body interaction includes the effects of image charges and
is given by:
\begin{equation}
g(\vec{r}_1,\vec{r}_2)=
\frac{1}{4\pi\epsilon}\,\frac{e^2}{|\vec{r}_1-\vec{r}_2|}-
\frac{1}{4\pi\epsilon}\,\frac{e^2}{\sqrt{|\vec{r}_1-\vec{r}_2|^2+(2z)^2}}-
\frac{1}{4\pi\epsilon}\,\frac{e^2}{2z}\,,
\label{twobody}
\end{equation} 
where $\epsilon=\epsilon_r \epsilon_0$ ($\epsilon_r$ and $\epsilon_0$ being
the relative permittivity of Gallium Arsenide and the vacuum permittivity, 
respectively), and $e$ is the electron charge. 
This expression has been obtained by taking one half of 
the electrostatic energy of the system made up of the 
electrons and their image counterparts, since the energy stored in the 
image space is purely
fictitious. The last term of Eq.~(\ref{twobody}), 
due to the interaction of each electron with
its own image, yields a constant shift of the energy spectrum.

To apply the CI method to the system described by the
Hamiltonian in Eq.~(\ref{2elH}), we start by
choosing a set of $n$ wave functions $\{\psi_i(\vec{r})\}$.
We shall refer to the $\{\psi_i(\vec{r})\}$'s as orbitals,
to distinguish them from the spin-orbitals $\{\varphi_i({\bf q})\}$.
By combining each of the $\psi_i$'s with one of the two possible spin
eigenstates
corresponding to the two spin orientations along the $z-$axis, we
obtain the set $\{\varphi_i({\bf q})\}$ of $M=2n$ spin-orbitals.

With $M=2n$ spin-orbitals and two electrons,
we can construct $n(2n-1)$ independent Slater
determinants (see Eq.~(\ref{nsd})). In the expansion of Eq.~(\ref{wf}), instead,
we take into account only $n^2$ Slater determinants, i.e., those
composed of spin-orbitals with opposite spins, that correspond to
states with zero total spin component $S_z$ along the $z-$axis.
These states are, in general, linear combinations of the singlet state
$|S=0,S_z=0>$ and of the triplet
state $|S=1,S_z=0>$. Since we are dealing with a spin-independent
Hamiltonian, the triplet states $|S=1,S_z=0>\,,|S=1,S_z=\pm 1>$ are 
degenerate in energy; in addition, the corresponding wave functions 
have the same spatial part. Therefore, no information about 
the energy eigenstates of the system is lost, if only one of the 
triplet states is used to expand the wave function of Eq.~(\ref{wf}). 

With the above mentioned restriction on the number of Slater
determinants, the matrix ${\cal H}$ is an $n^2\times n^2$ Hermitian
matrix. We can choose the orbitals $\psi_i$ to be real, thus
making the matrix ${\cal H}$ real and symmetric. Finally we notice that
${\cal H}$ is a full matrix, since the $\Phi_k$'s, being $2\times 2$ 
determinants, cannot differ by more than two spin-orbitals.

The choice of the set of orbitals  $\{\psi_i\}$ is of crucial
importance, since the number $n$ of orbitals required to get a
satisfactory approximation of the ground state energy and the 
corresponding wave
function depends on it. We have used as orbitals the
single-electron eigenstates for an isolated cell and we
have found that with this basis 12 orbitals (i.e. 24 spin orbitals) are
sufficient to
get a good accuracy in the results for cell sizes around 200~nm. In order 
to check the validity of
the approximation, we have also performed
calculations using 24 orbitals, finding  that the results are
practically identical to those obtained with the smaller basis.
For larger cell sizes, a larger number of orbitals would be necessary,
because the electrostatic interaction would grow in importance compared
to the confinement energy. Therefore, the ground state wave function
would deviate further from a single Slater determinant built with the
single-electron orbitals, and would thus need to be expanded on a basis of 
Slater determinants built from a larger set of single-electron 
wave functions.

Let us briefly comment on the relationship of the solution obtained with this
method to those obtained with other commonly
used approaches for the self-consistent solution of the electronic
structure of quantum dots. In the Hartree method\cite{Mac2} and in the
Local Density Functional Approximation (LDA)\cite{Mac1}, the exchange term
is ignored or treated in an approximate fashion, while the wave functions
are obtained at each iteration as eigenfunctions of a modified Hamiltonian.
When convergence is reached, the obtained wave functions are the set of
one-electron wave functions minimizing the expectation value of the
approximate Hamiltonian.

In the case of the Hartree-Fock method\cite{Pfann} the many-electron wave
function is represented by a single Slater determinant, and when 
self-consistency
is attained, the resulting Slater determinant minimizes the
expectation value of the Hamiltonian, as in the case of the Hartree and LDA
methods, but with a properly anti-symmetrized wave function.

With the Configuration-Interaction method, the wave function is
expanded over a basis of Slater determinants, which guarantees proper
antisymmetrization. The basis functions are fixed and the unknowns are
the coefficients of the expansion. If the basis were complete
(comprising an infinite number of spin-orbitals), the solution
would be exact and corresponding to that obtainable by diagonalizing
the many-body Hamiltonian.
In order to make the problem computationally feasible, we must limit the 
number of basis functions, introducing, as a result, some approximation. The 
difference between the solutions obtained with the application of the 
Hartree-Fock method and of the CI method can thus be summerized as 
follows: with the former we
get a single, optimized Slater determinant, while with the latter we obtain
an expansion of the solution over a finite basis of Slater determinants which
have been chosen ``a priori''. In the presence of strong electron-electron
correlation, the wave function obtained with the CI method with the inclusion 
of a reasonable number of basis functions is expected to be much closer to the 
exact solution than the optimized Slater determinant resulting from the 
Hartree-Fock method.

\section{Results and discussion}
\subsection{Single gate configuration with four circular holes}
We start the presentation of the numerical results with the cell-to-cell
response function obtained
for a cell defined by a gate with four circular holes, 
for a choice of 4, 12, and 24 basis orbitals (Fig.~\ref{fig6}).
Each cell is defined by holes
with a diameter of 90~nm, placed at the corners of a square, with a
distance of 110~nm between the hole centers. The applied gate bias is $-0.5$~V
and the separation between the centers of the two cells is 400~nm.
We found that there is no significant difference between the
results obtained with the three choices of basis elements. In some other
cases we have noticed a small difference between the response function
calculated with 4 basis orbitals and those for 12 and 24 orbitals, which were
instead found to be practically identical. All the results presented in the 
following 
have therefore been obtained with a basis of 12 orbitals.

An example of the electron density, computed for full polarization, is 
reported in Fig.~\ref{fig7}.

The distance of the 2DEG from the surface of the heterostructure plays
an important role and significantly affects the cell-to-cell response
function: the closer it lies to the surface, the higher, for a given
value of the bias voltage, are the potential barriers separating the
dots in a cell, and, as a result, the steeper and more abrupt is the response
function. This is the prevailing effect, even though it is partially
compensated by the screening action of the surface, which  reduces 
the cell-to-cell interaction and increases with
decreasing distance.

In Fig.~\ref{fig8}(a) we show a polarization curve which has been obtained 
for the same dot configuration as that in Fig.~\ref{fig6}, but for different
values of the 2DEG depth, from 45 to 55~nm. As expected, the polarization
curve becomes smoother for increasing depth of the 2DEG. The 
screening effect from the gates can be appreciated by comparing with 
the results obtained neglecting the contribution due to the images. 
This is seen in Fig.~\ref{fig8}(b), where we report the cell-to-cell 
response function for the previously considered gate geometry and for 
a 2DEG depth of 50~nm, computed with and without images. The image 
effects decrease the sharpness of the response function, because the 
dipole moment of the driver cell is screened by its images. 

As stated in the introduction, we have applied our
Configuration-Interaction technique to assess 
the sensitivity of a two-cell structure to fabrication tolerances. First 
we considered giving one of the holes in the gate a diameter
slightly different from the others: this leads to a variation in the potential
landscape defining the cell, and, in particular, to a modification
of the confinement energy associated with that dot. 
If we reduce the diameter of one of the holes,
the confinement energy for the dot underneath will rise by a certain
amount $\delta E$. If $\delta E$ is larger than the electrostatic
splitting $\Delta E_C$ between the two cell configurations, the cell will be
stuck in a state with the smaller dot empty.
For $\delta E < \Delta E_C$,
the cell will still be operational, but the cell-to-cell response
function will be shifted by an amount which depends on the
ratio of $\delta E$ to $\Delta E_C$. The strong nonlinearity of the
response function helps to restore the correct polarization value along
a chain of cells, as long as the shift still allows full
polarization of the driven cell for full polarization of the driver cell.

The tolerance on the hole diameter, admissible before unrecoverable
disruption of the operation of two coupled cells occurs, is unfortunately
very small:
from our calculations it is about one part in
10,000, for the dot sizes considered so far. Since the tolerance
is determined by the interplay between the electrostatic splitting
energy (which has an inverse linear dependence on the size) and the
perturbation of the confinement energy (which has an inverse quadratic
dependence on the size), it will be even tighter for smaller
cells.

In Fig.~\ref{fig9}(a) we report the cell-to-cell response function for a gate
configuration as previously described, with
the diameter of one of the dots reduced to 0.9999 times the nominal
value. The depth of the 2DEG is assumed to be 50~nm and the three solid 
curves refer
to different separations between cell centers. It is clear that, for a
separation between the centers greater than 280~nm, an error of one
part in 10,000 will be sufficient to unrecoverably disrupt QCA operation.
For purposes of comparison, we have reported also the results (dashed 
curves) obtained 
neglecting the contribution from the image charges (i.e. the screening 
due to the gates and to the assumed Fermi level pinning at the 
semiconductor-air interface).

We have also investigated the sensitivity to errors in the position 
of the gate holes. With a hard-wall model, such as in 
Ref.~\cite{Togsat}, a small shift in the position of one hole would 
not have a disrupting effect, since it would cause only a 
proportionately small variation of the electrostatic splitting and 
would not affect the confinement energy in any way. With a realistic 
model, as here, the situation is quite different: the confinement 
potential for each dot is determined not only by the corresponding 
hole, but also by the other holes belonging to the same cell. This 
means that by shifting a hole away from its nominal position, the 
potential landscape defining the dot underneath will be distorted, 
and the confinement energy will change. As a consequence, cell 
operation will be disrupted for unexpectedly small errors in hole 
positioning. In Fig.~\ref{fig9}(b) we report the cell-to-cell 
response function obtained for a 0.0275~nm shift down and to the 
left, of the bottom left dot of the driven cell. For the rest, the 
cell configuration is unchanged: dot diameter of 90~nm, dot-dot 
separation within a cell of 110~nm and distance between cell centers 
of 280~nm. Such a small displacement is sufficient to significantly 
shift the response function; a 0.05~nm shift leads already to the 
breakdown of cell operation. 

\subsection{Multiple independent gates with adjustable voltages} 
{}From the results shown above, it is clear that a simple hole-array 
implementation of QCA cells leads to unrealisable requirements on 
fabrication tolerances. This is the reason why we have also 
investigated alternative gate layouts such as that sketched in 
Fig.~\ref{fig4}. If dot confinement is obtained via multiple 
independent gates, it is indeed possible to compensate for 
geometrical tolerances by adjusting the gate voltages. 

In Fig.~\ref{fig10} we show the cell-to-cell response function for a 
cell defined with bias voltages of $-1.8$~V applied to all gates, 
except for gates 2 and 6, which are fixed at -1.6~V; these bias 
values have been chosen within a reasonable voltage range so as to 
get four clearly confined dots. The separation between the centers of 
the driver and the driven cells is 280~nm. Also in this case, 
different depths for the 2DEG have been considered, from 40 to 55~nm. 

A cross-section of the confinement potential, cut across
the two upper dots, is shown in Fig.~\ref{fig11} for a 2DEG at 35~nm 
(solid line), 45~nm (dotted line) and 55~nm (dashed line): the height 
of the barriers is relatively low, but their width guarantees a low 
enough transparency for strong localization of the electrons and 
thereby correct operation of the cell. 

We have also investigated the dependence of the cell-to-cell response 
function on electrical asymmetries, for the case of a 2DEG 50~nm 
depth and a distance of 280~nm between the centers of the driver and 
the driven cell. For this purpose, the voltage applied to gate 3 has 
been made slightly more negative by an amount $\delta V$. The results 
for $\delta V= -0.05, -0.1, -0.2$~mV are shown in 
Fig.~\ref{fig12}(a). The most visible effect is a shift in the 
cell-to-cell response function, which is somewhat proportional to the 
variation in the applied voltage. This does not disrupt the operation 
of a QCA chain, as long as full polarization of one cell can produce 
full polarization of the neighboring cell. Therefore, for this 
particular cell, we expect a maximum tolerance on the gate voltages 
of about 0.4~mV. This may seem difficult to achieve at first sight, 
but it is important to keep in mind that it is a shift between gate 
biases: larger variations in the overall average value of the gate 
voltages are allowed, as long as they do not alter cell occupancy. 

As previously mentioned, the screening effect due to the gates and to 
the charge at the semiconductor-air interface decreases the strength 
of the electrostatic interaction and, therefore, the energy splitting 
between the two possible cell polarizations. Hence, the effects of 
asymmetry decrease with increasing depth of the 2DEG, as the image 
effects are reduced. This phenomenon is clearly visible in 
Fig.~\ref{fig12}(b), where the cell-to-cell response function is 
plotted for a voltage shift of $-0.1$~mV on gate 3 and various values 
of the 2DEG depth ranging from 35~nm to 50~nm. The shift in the 
response function decreases as the depth of the 2DEG increases, and 
the effect of the image term becomes less important. 

Fabrication tolerances would also disrupt the operation of this type of 
cell, but they can be compensated for by fine adjustments to 
the bias voltages applied at the gates. As an example, 
we consider a cell with gate 3 shifted by 5~nm to the right. An 
iterative procedure was developed for computing the new 
bias voltages that will restore the symmetry of the structure. For a 
perfectly symmetric structure there is a four-fold quasi-degeneracy 
of the one-electron states, corresponding to the four-fold symmetry 
of the cell. When symmetry is disrupted, this quasi-degeneracy is 
lifted, and the first four eigenvalues differ from each other by an 
amount which is no longer negligible. Our strategy is to evaluate the 
difference between the first and the fourth eigenvalue, and then 
adjust each gate voltage in such a way as to minimize this 
difference. We vary one gate voltage 
at a time, from gate 1 to gate 7 (with the exception of gate 4, which does not 
affect cell symmetry), and then repeat the cycle, starting 
again from gate 1, until the splitting between the first and the 
fourth eigenvalue is smaller than an assigned threshold. In order to 
avoid getting stuck in local minima, it is convenient to perform the 
whole minimization procedure several times, for values of the 
displacement of gate 3 increasing with a geometric progression from 0.01~nm 
to 5~nm. 
Finally, small adjustments are made 
manually, until a symmetric cell-to-cell response function is 
obtained. The gate voltages needed to symmetrize the cell are 
listed in Table~\ref{tab1}, while the cell response function for 
the symmetrized cell is shown in Fig.~\ref{fig13}. This demonstrates 
that a 10\% error in the position of one of the gates can be fully 
compensated. State of the art fabrication techniques allow 
geometrical tolerances of this order of magnitude or smaller, 
thus such a cell is actually manufacturable, although not useful 
for large scale applications where it would be impossible to tune 
each cell separately. By acting on the gate voltages it is also 
possible to compensate for the presence of randomly distributed stray 
charges, which would also disrupt QCA operation. We notice that the 
cell-to-cell response function of Fig.~\ref{fig13} is steeper than that
for a geometrically symmetric cell with analogous parameters (see 
Fig.~\ref{fig10}). This is a consequence of a slight variation in the barrier
heights and widths (due to the different applied voltages) and of the 
exponential dependence of the tunneling coefficients between the dots on 
such barrier characteristics.
A chain of such cells can be fabricated by repeating this same gate layout
in the horizontal direction. However, lateral branching, needed 
for the implementation of logic gates\cite{Toglog}, is not allowed, 
due to the lateral extension of the leads required for feeding the bias 
voltages. From this point of view, a more promising implementation would
be that suggested by Chen and Porod\cite{Porochen}, with central enhancement 
gates in each dot: by adjusting the voltage of such a central gate it would
be possible to correct asymmetries, while keeping the possibility of 
lateral branching. This implementation, however, poses serious fabrication
problems, because of the difficulties involved in separately contacting
all of the central gates.

We emphasize that the compensation procedure described above is not 
proposed as a practical method, rather our aim has been to 
demonstrate that compensation is possible in principle.

\section{Conclusions}
We have shown that the Configuration Interaction method allows us to
solve the many-electron Schr\"odinger equation for a QCA cell
made up of four coupled dots. This represents
a definite advance over the state of the art in the simulation of multiple
quantum dot systems: it is possible to include realistic confinement 
potentials without resorting to some type of
mean-field approximation (local density approximation, Hartree-Fock,
Hartree-Fock-Roothaan), which fail to converge when single-electron states are
strongly degenerate and the electrostatic interaction is comparable to
the confinement energy.
A Hubbard-like approach to QCA cells is also feasible
and has been shown to provide a qualitative understanding of the underlying 
physics\cite{Togsat},
but requires a set of phenomenological parameters, such as the
on-site electrostatic interaction, the dot confinement energy and the
tunneling energy, which cannot be easily obtained from the geometrical
structure and from experiments. With the CI method it is sufficient
to determine the confinement potential from the 
layer structure and the gate layout.
This is an important advantage, which makes simulations
based on the CI method a reliable and effective design tool.

A disadvantage of the CI approach lies in the large computational 
resources which are required, since the number of Slater determinants 
to be considered exhibits a combinatorial increase with the number of 
electrons in the system. In this paper, we have presented results for 
two electrons, and a QCA cell with up to six electrons is currently 
being investigated. Calculations for a larger number of electrons 
would require prohibitive memory sizes (well above 1 Gbyte) or 
extremely long computation times. 

We have focused our investigation on the sensitivity to fabrication 
tolerances for two coupled QCA cells. Our results demonstrate that 
the implementation of a simple ``hole-array'' approach is not 
feasible, because it would require a precision in the diameter of 
each hole that is well beyond the present state of the art in 
electron-beam lithography. 

We have proposed an alternative cell layout, based on seven gates, whose
bias voltages can be independently adjusted: this approach is within the 
capabilities of current fabrication technologies, since geometrical 
errors in the gate positions can be corrected by means of appropriate voltage 
variations.
Admissible voltage tolerances are rather small, but achievable using 
resistive voltage dividers cooled down together with the sample. 
The approach we are proposing should allow an experimental demonstration
of the QCA principle, from a single cell up to a chain of cells, but it is 
not suitable for the realization of logic gates, due to the impossibility of 
lateral branching, which is prevented by the leads reaching each gate. 
It is also impractical 
for large-scale integration, due to the need for individual adjustment of
each single cell. Viable logic circuits will require drastically
different solutions and new architectural concepts.

\acknowledgments
We thank Prof. C. Guidotti for her useful suggestions on the CI method
and Prof. D. W. L. Sprung for valuable discussion.
This work has been supported by the ESPRIT Project 23362 QUADRANT
(QUAntum Devices foR Advanced Nano-electronic Technology).

\appendix
\section{}

The problem of computing the matrix elements of the
Hamiltonian in Eq.~(\ref{Htot}) between two Slater determinants is well
known~\cite{Slater}. For the diagonal elements one finds:
\begin{eqnarray}
&&\langle \Phi_k|\hat{H}|\Phi_k\rangle=
\sum_{i}\langle \varphi_{n_{ik}}|h|\varphi_{n_{ik}}\rangle\nonumber \\
&&+\frac{1}{2}\,
\sum_{ij}(\langle \varphi_{n_{ik}}\varphi_{n_{jk}}|g|\varphi_{n_{ik}}
\varphi_{n_{jk}}\rangle -
\langle \varphi_{n_{ik}}\varphi_{n_{jk}}|g|\varphi_{n_{jk}}
\varphi_{n_{ik}}\rangle)\,,
\label{Hd}
\end{eqnarray}
where in general
\begin{equation}
\langle \varphi_{i}\varphi_j|g|\varphi_l\varphi_m\rangle =
\int\,dq_1\,dq_2\,\varphi_{i}^{*}(q_1)\varphi_j^{*}(q_2)
g(\vec{r}_1,\vec{r}_2)
\varphi_l(q_1)\varphi_m(q_2)\,.
\end{equation}

As far as the computation of the off-diagonal matrix elements of
the Hamiltonian of Eq.~(\ref{Htot}) between two different Slater
determinants $\Phi_k\,,\Phi_{k^\prime}$ is concerned, there are some
``selection rules'' (Slater's rules~\cite{Lverde})
which state that there are only two possible cases in which
$\langle \Phi_k |\hat{H}|\Phi_{k^\prime}\rangle$ is not vanishing, i.e. when
$\Phi_k\,,\Phi_{k^\prime}$ either differ by one single spin-orbital or by
two:
\begin{enumerate}
\item {\sl one spin-orbital difference}
($\varphi_{n_{ik}}\ne \varphi_{n_{ik^\prime}}$)
\begin{eqnarray}
&&\langle \Phi_k|\hat{H}|\Phi_{k^\prime}\rangle=
\langle\varphi_{n_{ik}}|h|\varphi_{n_{ik^\prime}}\rangle+\nonumber \\
&&
\sum_{j\ne i}(\langle
\varphi_{n_{ik}}\varphi_{n_{jk}}|g|\varphi_{n_{ik^\prime}}
\varphi_{n_{jk}}\rangle -
\langle \varphi_{n_{ik}}\varphi_{n_{jk}}|g|\varphi_{n_{jk}}
\varphi_{n_{ik^{\prime}}}\rangle)\,,
\label{Hnd1}
\end{eqnarray}
\item {\sl two spin-orbital difference}
($\varphi_{n_{ik}}\ne \varphi_{n_{ik^\prime}}$ and
$\varphi_{n_{jk}}\ne \varphi_{n_{jk^\prime}}$ )

\begin{equation}
\langle \Phi_k|\hat{H}|\Phi_{k^\prime}\rangle=
\langle
\varphi_{n_{ik}}\varphi_{n_{jk}}|g|\varphi_{n_{ik^\prime}}
\varphi_{n_{jk^\prime}}\rangle -
\langle \varphi_{n_{ik}}\varphi_{n_{jk}}|g|\varphi_{n_{jk^\prime}}
\varphi_{n_{ik^{\prime}}}\rangle\,.
\label{Hnd2}
\end{equation}
\end{enumerate}

The expressions in Eqs.~(\ref{Hnd1}, \ref{Hnd2}) refer to the case in which the
spin-orbitals that are common to both Slater determinants occur in
the same columns. If this is not the case, it is possible to perform a
permutation of the columns of one determinant, so that the above condition
is satisfied; the permutation has the effect of changing the sign of
the matrix element if it is an odd order permutation.

Finally, it is worth mentioning that Eqs.~(\ref{Hd}, \ref{Hnd1}, \ref{Hnd2})
are valid only if the orthonormality condition on the spin-orbitals
is satisfied.

\begin{figure}
\caption {Schematic representation of two coupled QCA cells: tunneling
of electrons is possible along the dashed lines.}
\label{fig1}
\end{figure}

\begin{figure}
\caption {Subdivision of a cell into four quadrants.}
\label{fig2}
\end{figure}

\begin{figure} 
\caption {QCA cell obtained by depleting a two-dimensional electron
gas by means of a metallic gate with four holes, which define the four
quantum dots.}
\label{fig3}
\end{figure}

\begin{figure}
\caption {Gate layout for the definition of four independently 
adjustable quantum dots; all distances are in nanometers.}
\label{fig4}
\end{figure}

\begin{figure} 
\caption {Confinement potential at a depth of 50~nm from the surface of the 
heterostructure, produced by a gate with four holes of diameter 90~nm
and placed at the corners of a 110~nm square.}
\label{fig5}
\end{figure}

\begin{figure}
\caption {Cell-to-cell response function obtained for a cell defined by 
a gate, kept at $-0.5$~V, with four 90~nm holes placed at the corners of 
a 110~nm square.
The depth of the 2DEG is 50~nm and the separation between the centers
of the driven and the driver cells is 400~nm. The different symbols 
indicate different numbers of basis functions.}
\label{fig6}
\end{figure}

\begin{figure}
\caption {Electron density, in arbitrary units, for a completely 
polarized cell defined by a gate, kept at $-0.5$~V, with four 90~nm holes 
placed at the corners of a 110~nm square. The depth of the 2DEG is 50~nm.}
\label{fig7}
\end{figure}

\begin{figure}
\caption {Cell-to-cell response functions obtained for a cell defined by 
a gate, kept at $-0.5$~V, with four 90~nm holes placed at the corners of a 
110~nm square.
The separation between the centers of the two cells is 400~nm. (a) Comparison
between the response functions for different values of the 2DEG depth. (b)
Comparison between the response functions obtained with and without the
contribution of the image charges.}
\label{fig8}
\end{figure}

\begin{figure}
\caption {Cell-to-cell response functions obtained for a cell defined by 
a gate, kept at $-0.5$~V, with four holes placed at the corners of a 
110~nm square. (a) The bottom left hole has a diameter of 89.991~nm and the
other 3 holes of 90~nm; the depth of the 2DEG is 50~nm. Results for three 
different values of the separation between the cell centers are reported,
including the effect of image charges (solid lines) or neglecting it (dashed
lines). (b) The four holes have a diameter of 90~nm, but the bottom left hole 
has been moved down and to the left by 0.0275~nm.}
\label{fig9}
\end{figure}

\begin{figure}
\caption {Cell-to-cell response functions obtained for a cell defined 
with the seven-gate layout 
and a distance between 
cell centers of 280~nm. The gates are kept at 
$-1.8$~V, except for gates 2 and 6, which are kept at $-1.6$~V. 
Results for a depth of the 2DEG varying between 40 and 55~nm have
been reported.}
\label{fig10}
\end{figure}

\begin{figure}
\caption {Cross-section of the confining potential obtained with the
seven-gate layout, cut across the two upper dots, at a depth from the surface 
of 35~nm (solid line), 45~nm (dotted line), and 55~nm (dashed line).}
\label{fig11}
\end{figure}

\begin{figure}
\caption {Cell-to-cell response functions obtained for a cell defined 
with the seven-gate layout 
and for a distance between 
cell centers of 280~nm. The gates are kept at 
$-1.8$~V, except for gates 2 and 6, which are kept at $-1.6$~V, and
gate 3, which is kept at $-1.8$~V$+\delta V$. (a) The depth of the 2DEG
is 50~nm and the results for three different values of $\delta V$ have
been reported. (b) The value of $\delta V$ is $-0.1$~mV and results for 
different depths of the 2DEG are shown.}
\label{fig12}
\end{figure}

\begin{figure}
\caption {Cell-to-cell response functions obtained for a cell defined
with the seven-gate layout, 
a distance between
cell centers of 280~nm, and a 2DEG depth of 50~nm. Gate 3 has been shifted 
to the right by 5~nm, and the bias voltages are those listed in Table~1, 
chosen to restore cell symmetry.}
\label{fig13}
\end{figure}

\newpage

\narrowtext
\begin{table}
\caption{Values of the bias voltages to be applied to the 
gates defining the geometrically asymmetric cell shown in Fig.~4,
in order to symmetrize it.}
\begin{tabular}{cc}
Gate &  Bias voltage\\
\tableline
$V_1$ & $-1.665577$ V \\
$V_2$ & $-1.728683$ V \\
$V_3$ & $-1.845800$ V \\
$V_4$ & $-1.800000$ V \\
$V_5$ & $-1.807502$ V \\
$V_6$ & $-1.592665$ V \\
$V_7$ & $-1.793715$ V \\
\end{tabular}
\label{tab1}
\end{table}

\end{document}